\documentclass[prl,aps,superscriptaddress,twocolumn, amsmath]{revtex4-1}
\usepackage{graphicx}
\usepackage{color}
\usepackage{multirow}
\usepackage{hhline}
\usepackage{siunitx}

\newcommand {\ket}[1] {|{#1}\rangle}

\begin{document}

\title{Deterministic photon-emitter coupling in chiral photonic circuits}

\author{Immo S\"{o}llner} \email{sollner@nbi.ku.dk}
\affiliation{Niels Bohr Institute, University of Copenhagen, Blegdamsvej 17, DK-2100 Copenhagen, Denmark}
\author{Sahand Mahmoodian}
\affiliation{Niels Bohr Institute, University of Copenhagen, Blegdamsvej 17, DK-2100 Copenhagen, Denmark}
\author{Sofie Lindskov Hansen}
\affiliation{Niels Bohr Institute, University of Copenhagen, Blegdamsvej 17, DK-2100 Copenhagen, Denmark}
\author{Leonardo Midolo}
\affiliation{Niels Bohr Institute, University of Copenhagen, Blegdamsvej 17, DK-2100 Copenhagen, Denmark}
\author{Alisa Javadi}
\affiliation{Niels Bohr Institute, University of Copenhagen, Blegdamsvej 17, DK-2100 Copenhagen, Denmark}
\author{Gabija Kir{\v{s}}ansk{\.{e}}}
\affiliation{Niels Bohr Institute, University of Copenhagen, Blegdamsvej 17, DK-2100 Copenhagen, Denmark}
\author{Tommaso Pregnolato}
\affiliation{Niels Bohr Institute, University of Copenhagen, Blegdamsvej 17, DK-2100 Copenhagen, Denmark}
\author{Haitham El-Ella}
\affiliation{Niels Bohr Institute, University of Copenhagen, Blegdamsvej 17, DK-2100 Copenhagen, Denmark}
\author{Eun Hye Lee}
\affiliation{Center for Opto-Electronic Convergence Systems, Korea Institute of Science and Technology, Seoul, 136-791, Korea}
\author{Jin Dong Song}
\affiliation{Center for Opto-Electronic Convergence Systems, Korea Institute of Science and Technology, Seoul, 136-791, Korea}
\author{S{\o}ren Stobbe}
\affiliation{Niels Bohr Institute, University of Copenhagen, Blegdamsvej 17, DK-2100 Copenhagen, Denmark}
\author{Peter Lodahl} \email{lodahl@nbi.ku.dk} \homepage{www.quantum-photonics.dk}
\affiliation{Niels Bohr Institute, University of Copenhagen, Blegdamsvej 17, DK-2100 Copenhagen, Denmark}

\date{\today}

\maketitle

{ \bf{The ability to engineer photon emission and photon scattering is at the heart of modern photonics applications ranging from light harvesting, through novel compact light sources, to quantum-information processing based on single photons. Nanophotonic waveguides are particularly well suited for such applications since they confine photon propagation to a 1D geometry thereby increasing the interaction between light and matter. Adding chiral functionalities to nanophotonic waveguides lead to new opportunities enabling integrated and robust quantum-photonic devices or the observation of novel topological photonic states. In a regular waveguide, a quantum emitter radiates photons in either of two directions, and photon emission and absorption are reverse processes. This symmetry is violated in nanophotonic structures where a non-transversal local electric field implies that both photon emission \cite{Mitsch2014arXiv, Luxmoore2013PRL} and scattering \cite{Junge2013PRL} may become directional. Here we experimentally demonstrate that the internal state of a quantum
emitter determines the chirality of single-photon emission in a specially engineered photonic-crystal waveguide. Single-photon emission into the waveguide with a directionality of more than 90\% is observed under conditions where practically all emitted photons are coupled to the waveguide. Such deterministic and highly directional photon emission enables on-chip optical diodes, circulators operating at the single-photon level, and deterministic quantum gates. Based on our experimental demonstration, we propose an experimentally achievable and fully scalable deterministic photon-photon CNOT gate, which so far has been missing in photonic quantum-information processing where most gates are probabilistic \cite{Obrien2009NPHOT}. Chiral photonic circuits will enable dissipative preparation of entangled states of multiple emitters \cite{Stannigel2012NJP}, may lead to novel topological photon states \cite{Hafezi2011NPHYS, Kraus2012PRL}, or can be applied in a classical regime to obtain highly directional photon scattering \cite{Rodriguez2013Science, Petersen2014Science, Neugebauer2014NL}}. }

\begin{figure*}[t!]
	\includegraphics[width=\textwidth]{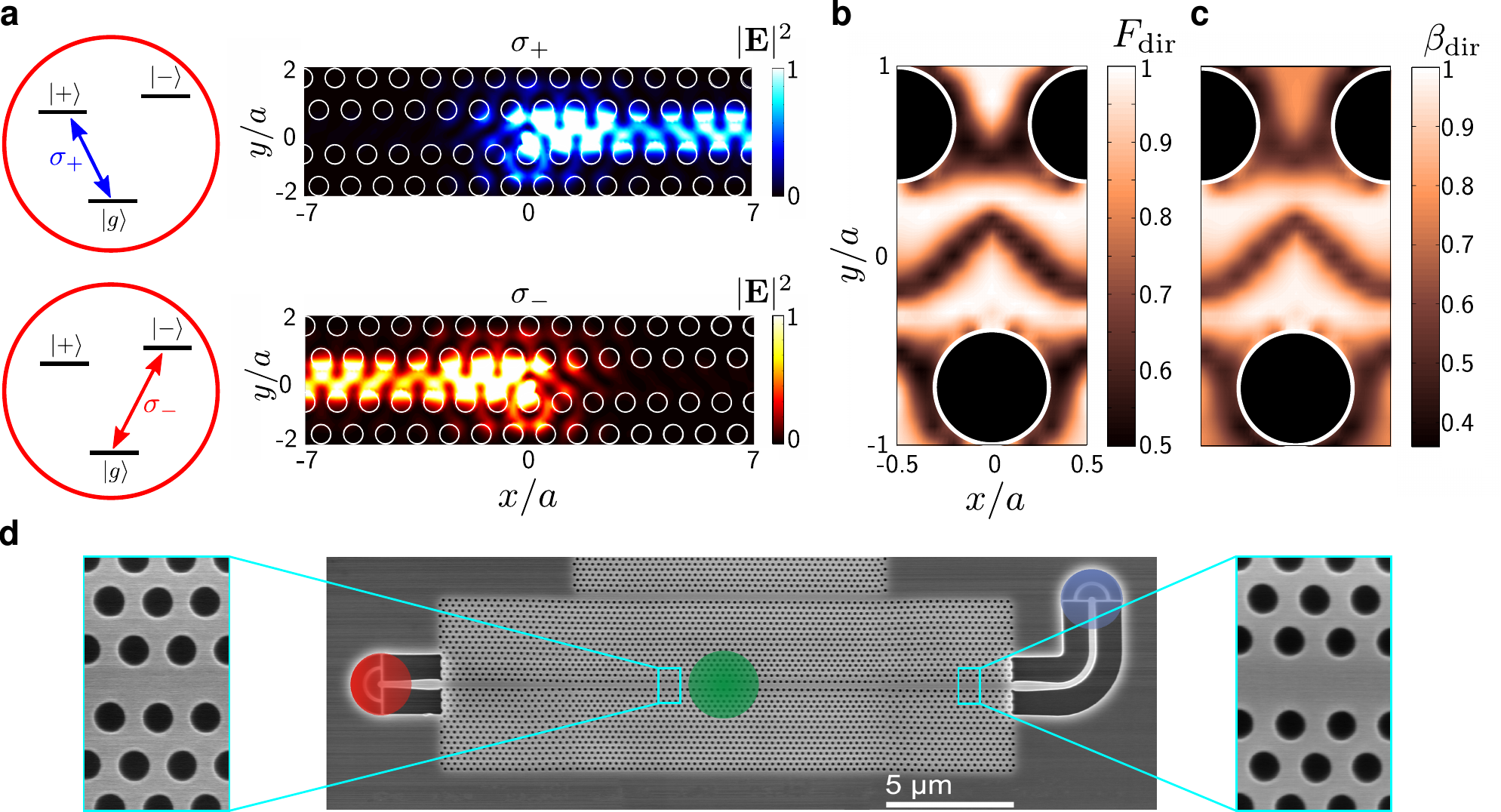}
	\caption{ \label{fig:fig1} Operational principle of chiral single-photon emission in a waveguide. {\bf a)} Left: QD level scheme in the presence of a magnetic field in the growth directions featuring two circularly polarized exciton transitions $\sigma_{\pm}$ with a splitting controlled by the magnetic field. Right: calculated directional emission patterns of $\sigma_{+}$ and $\sigma_{-}$ polarized dipole emitters in GPWs. {\bf b)} Directionality and {\bf c)} directional $\beta$-factor as a function of position for the GPW mode studied in the experiments. $a$ is the lattice constant of the photonic crystal. {\bf d)} Scanning-electron microscope image of the GPW. A QD is excited in the center of the structure (green area) and photons are collected from the out-coupling gratings at each side of the GPW (red and blue areas). The zoom-ins show the structure of the photonic-crystal lattice that consists of a GPW section (left image) and a standard photonic-crystal waveguide section (right image). The waveguide gradually changes from glide-plane lattice to a standard lattice in order to obtain efficient out coupling of photons. For details, see Supplementary Material.  }
\end{figure*}

Truly 1D photon-emitter interfaces are desirable for a range of applications in photonic quantum-information processing \cite{Lodahl2013arXivRMP}. To this end, photonic-crystal waveguides constitute an ideal platform featuring on-chip integration with the ability to engineer the light-matter coupling. Recent experiments have achieved a coupling efficiency for a single quantum dot (QD) to a photonic-crystal waveguide in excess of 98\%, thus constituting a deterministic 1D photon-emitter interface \cite{Arcari2014PRL}. Standard photonic-crystal waveguides are  mirror symmetric around the center of the waveguide and as a consequence the mode polarization is predominantly linear at the positions where light intensity is high. By designing a photonic-crystal waveguide that breaks this symmetry, modes that are circularly polarized at the field maxima can be engineered. We refer to this novel type of waveguide as a glide-plane waveguide (GPW), cf. Supplementary Material for further descriptions of the structural parameters. In a GPW, a QD with a circularly polarized transition dipole emits preferentially into a single direction as determined by the helicity, see Fig.~\ref{fig:fig1}.a. The exact branching ratio between the two emission directions is controlled by structural engineering and depends on the position of the QD. The two figures-of-merit for chiral single-photon emission are displayed in Figs. \ref{fig:fig1}.b and \ref{fig:fig1}.c. They are the fraction of photons emitted into the waveguide that propagates in a desired direction $(F_{\mathrm{dir}})$ and the overall probability that an emitted photon from the QD is channeled into the correct direction $(\beta_{\mathrm{dir}})$. The GPW is engineered to increase the $\beta$-factor and features $\beta_{\mathrm{dir}} = 98 \%$, cf. Fig.~\ref{fig:fig1}.c, which enables the deterministic interfacing of single emitters and single photons that is not possible in other waveguide systems \cite{Mitsch2014arXiv, Luxmoore2013PRL}. An ideal  directional photon-emitter interface corresponds to $\beta_{\mathrm{dir}} \rightarrow 1$, which likely can be achieved by further engineering of the GPW. We emphasize the robustness of the directional coupling in the sense that the GPW is designed to have highly directional coupling at many positions within the unit cell of the photonic-crystal waveguide.
The present work reports on the observation of highly directional emission at the single-photon level by efficiently coupling a single QD to a GPW. We show that this is the basic operational principle required for constructing a deterministic CNOT gate for photons.

Figure \ref{fig:fig1}.d displays a scanning electron microscope graph of our photonic waveguide and illustrates the directional coupling.  Single self-assembled QDs in the GPW are excited optically and two non-degenerate circularly polarized exciton states $\ket{+}$ and $\ket{-}$ (cf. Fig.~\ref{fig:fig1}.a) are formed by applying a strong magnetic field ($B_z$) in the QD growth direction \cite{Bayer2002PRBb}. By non-resonant optical excitation, a statistical mixture of the two exciton states are prepared that can decay to the ground state via $\Delta m= \pm 1$ dipole transitions emitting $\sigma_{\pm}$ polarized photons. These QD transitions are used to demonstrate chiral photon emission. The origin of the chiral interaction can be understood as follows: from time-reversal symmetry, the electric fields of counter-propagating modes with wave vectors ${\mathbf k}$, and $-\mathbf{k}$ satisfy ${\mathbf E}_{-{\mathbf k}}(\mathbf r) = {\mathbf E}_{\mathbf k}^*(\mathbf r)$, i.e., for a mode with an in-plane circular polarization, the counter-propagating mode has the orthogonal circular polarization. Therefore, a circularly polarized emitter only couples to the mode that has the same circular polarization as the dipole transition, which leads to unidirectional emission.

An important advantage of photonic crystals is that they allow tailoring the interaction between light and matter. In a GPW, the chirality is engineered by the structural parameters and depends on the position of the QD, as determined by the magnitude of the projection of the local electric field onto the QD transition dipole moment. 
The experimental proof of directional emission is obtained by collecting single photons emitted from the QDs by two separate outcoupling gratings at each end of the GPW, cf. Fig.~\ref{fig:fig1}.d. We extract the directionality factor $F_{\textrm{dir}}$ by comparing the intensity of one circular dipole to the orthogonal dipole by collecting the intensity from the same waveguide ends, which makes the method insensitive to potentially different outcoupling efficiencies from the two ends. By spectrally resolving the emission, single QD lines are selected and photon correlation measurements are employed in order to identify separate QDs and quantify the single-photon nature of the emission.

\begin{figure*}[t!]
	\includegraphics[width=\textwidth]{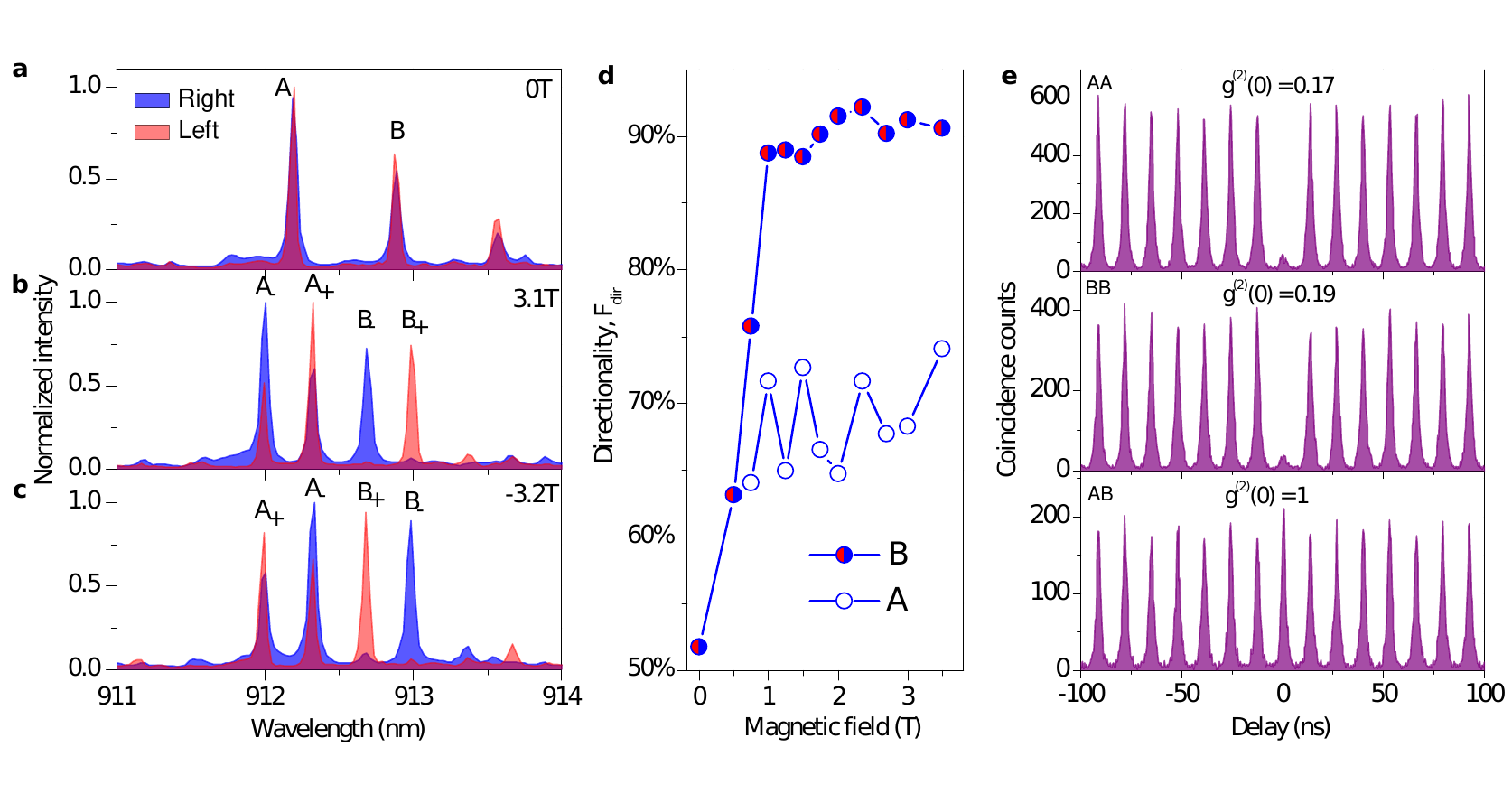}
	\caption{ \label{fig:fig2} Observation of directional emission of single QDs in a GPW. {\bf a)} Emission spectra displaying two different QD lines denoted A and B that are recorded on the right (blue spectrum) or the left (red spectrum) grating outcoupler. {\bf b)} By applying a magnetic field the QD lines split in duplets $\textrm{A}_{\pm}$ and $\textrm{B}_{\pm}$ that display different directionality. {\bf c)} Emission spectrum recorded for a negative magnetic field. For the opposite polarity the two exciton lines swap spectral position while the directional outcoupling is preserved. {\bf d)} Extracted directionality for QD A and B as a function of applied magnetic field. At low magnetic fields the lines $\textrm{B}_-$ and $\textrm{B}_+$ are not sufficiently separated, which leads to a systematic underestimate of the directionality factor. At approximately 1~T the directionality levels off and stays constant for the entire range of magnetic fields investigated. An average directionality factor of $F_{\textrm{dir}}=90\%\pm1.3\%$ is obtained for QD B. {\bf e)} Demonstration of single-photon operation of the chiral waveguide. AA (BB) denotes correlation measurements on QD line A (B) coupled out from the two separate ends of the GPW. The single-photon purity of the emission is quantified by extracting $g^{(2)}(0)$ from the data \cite{SantoriBook} where $g^{(2)}(0) < 1/2$ is the distinct anti-bunching signature of single-photon emission. AB denotes a cross correlation between the two QD lines A and B. The lack of any correlations prove that A and B correspond to two independent QDs. The data are taken at 0~T.}
\end{figure*}

The directionality of the photon emission is extracted from emission spectra measured for different applied magnetic field strengths, cf. Fig.~\ref{fig:fig2}.a-c  and two QD lines (A and B) are studied in detail. Without a magnetic field, the spectra recorded from the two ends of the GPW are almost identical. By increasing the magnetic field, the individual QD lines split into pairs corresponding to the two circularly polarized transitions.  Highly directional emission is evident for the peaks labelled $B_+$ and $B_-$. Furthermore, Figs. \ref{fig:fig2}.b and \ref{fig:fig2}.c show that the two transitions maintain their directionality when changing the polarity of the magnetic field, which swaps the spectral position of orthogonally polarized emission lines. This demonstrates explicitly that the directionality is related to the helicity of the emitted photon.

We plot the directionality of the two QD lines as a function of applied magnetic field strength in Fig.~\ref{fig:fig2}.d. At low magnetic fields the emission lines are not clearly separated, which leads to a systematic underestimation of the directionality factor. At approximately 1~T the directionality levels off and we extract $F_{\textrm{dir}}=90\%\pm1.3\%$ for QD B by averaging over the plateau region in Fig.~\ref{fig:fig2}.d. The lower directionality of QD A stems from the spatial variation within the unit cell of the GPW, cf. Fig.~\ref{fig:fig1}.b. It should be emphasized that the extracted directionality constitutes a lower bound of the actual value due to the presence of weak emission from transitions other than the investigated QD, which is due to the non-resonant excitation method applied in the experiment. Consequently, the actual directionality for the single QD transition is likely to approach unity in accordance with theory (Fig.~\ref{fig:fig1}.b). The single-photon nature of the emission can be proven through correlation measurements, see Fig.~\ref{fig:fig2}.e. A pronounced anti-bunching is observed for peaks A and B illustrating that high-purity single-photon emission is observed. Furthermore, the absence of correlations in the cross-correlation measurement between A and B shows that the two peaks originate from two independent QDs.

A major asset of the photonic-crystal waveguide platform is that light-matter interaction can be controlled to such a degree that the photon-emitter interface becomes deterministic \cite{Lodahl2013arXivRMP}, as is quantified by the $\beta$-factor. In time-resolved measurements, we record a decay rate of $ 0.80 \pm 0.02 \textrm{ns}^{-1}$ for QD B, which restricts the possible spatial position of the QD as well as its spectral position relative to the photonic waveguide bands of the GPW. From numerical calculations, we estimate an upper bound on the rate at which the QD leaks to non-guided waveguide modes, and by accounting also for the contribution from intrinsic non-radiative decay processes of the QD \cite{Arcari2014PRL}, we arrive at $\beta \gtrsim 90 \%.$  The detailed measurements of the $\beta$-factor in standard photonic-crystal waveguides were presented in Ref. \cite{Arcari2014PRL}.

\begin{figure*}[t!]
	\includegraphics[width=0.8 \textwidth]{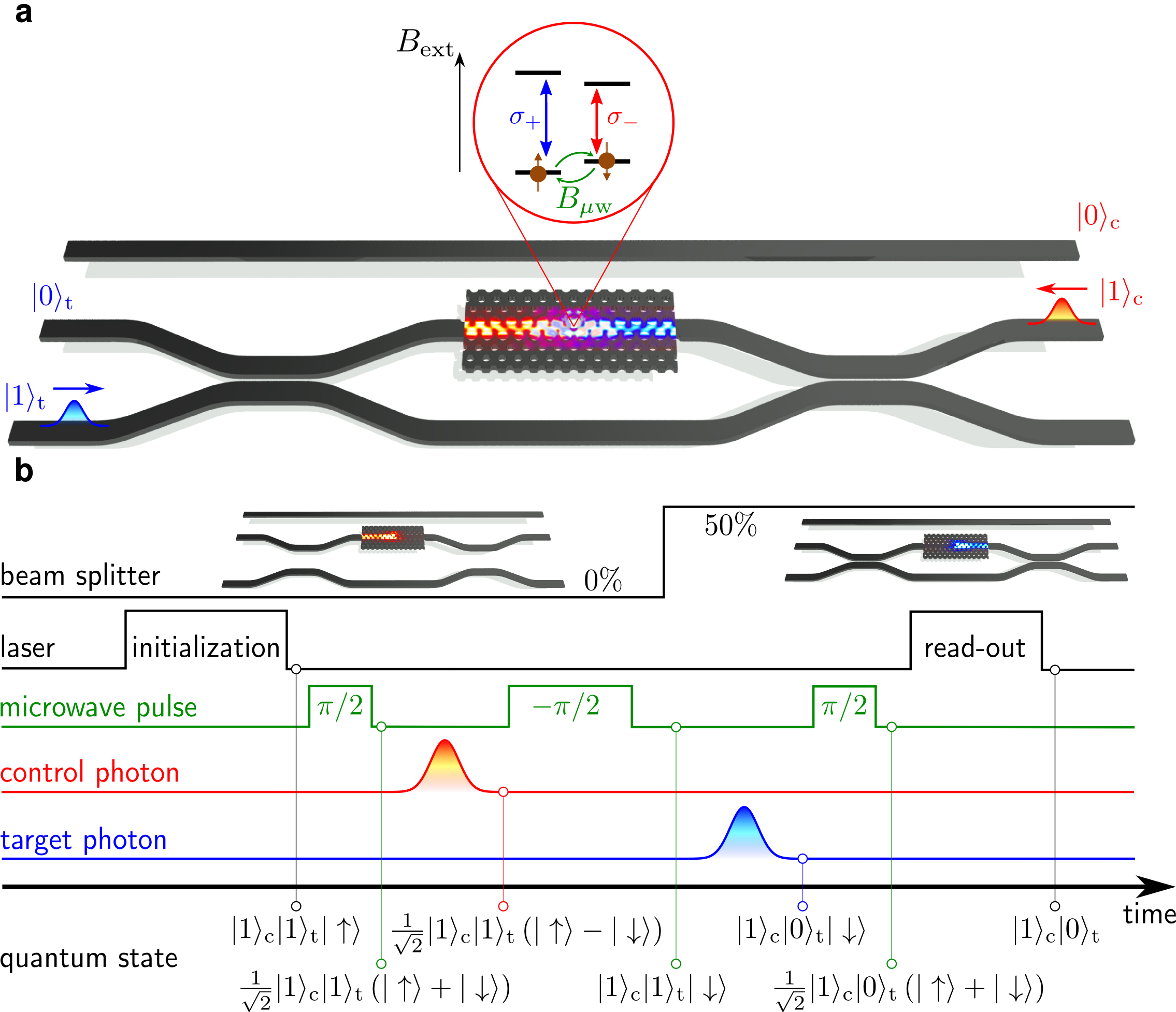}
	\caption{ \label{fig:fig3} Deterministic CNOT gate for photons based on the chiral spin-photon interface. {\bf a)} Layout of the gate consisting of three access waveguides where a control photon (red) can be injected from the right in a superposition of the two upper waveguides and controls the target photon (blue) that is injected from the left in the two lower waveguides. A single QD is embedded in the chiral GPW section and a single spin state coherently prepared with microwave pulses. The associated 4-level scheme and relevant circularly polarized optical transitions are indicated. The waveguide circuit is reconfigurable such that the three waveguides are either independent (initial setting) or the two lower waveguides are tuned in order to form a balanced Mach-Zehnder interferometer by inducing two 50/50 beamsplitters. {\bf b)} Operation sequence of the CNOT gate. Resonant laser pulses are employed to initialize and read out the spin state of the QD. The spin state of the QD is prepared in superposition states with $\pi/2$ rotation pulses (with, e.g., microwave pulses $B_{\mu W}$). The control photon and the first two microwave pulses impart a spin flip on the QD if the control photon is in state $\left| 1 \right>_{\mathrm c}$. Subsequently the target photon is directed to either $\left| 0 \right>_{\mathrm t}$ or $\left| 1 \right>_{\mathrm t}$ determined by the state of the control photon. The last $\pi/2$ rotation and subsequent spin measurement erases the residual quantum information in the spins. The evolution of the quantum state of the system for a specific initial condition is shown in the bottom of the figure.   }
\end{figure*}

Having demonstrated the basic operational principle of deterministic chiral photon emission, this functionality can be exploited for the construction of an experimentally feasible architecture of a deterministic CNOT gate for photons, which is the non-trivial two-photon gate enabling quantum computing \cite{Nielsen2010Book}. Figure  \ \ref{fig:fig3} displays the layout of the gate based on the chiral spin-photon interface where qubits are encoded in the photon path, i.e., the propagation of single photons in a waveguide. The required interaction between light and matter is mediated by the scattering of single photons on single emitters \cite{Duan2004PRL, Chang2008NPHYS, Shen2005OL}. The basic idea is that a control photon deterministically changes the internal state of the QD that can be prepared in a superposition state of an electron (or hole) being either spin up or down. The level scheme for the relevant trion QD states is shown in Fig.~\ref{fig:fig3}.a. Subsequently a $\pi$ phase shift \footnote{We note that the acquired transmission coefficient is at variance with the results reported in \cite{Shen2011PRL}. This mistake has been communicated to and acknowledged by the authors of the mentioned reference.} is induced on a counterpropagating target photon in the event that the control photon did interact with the QD. Before the arrival of the target photon, the optical circuit is reconfigured to a balanced Mach-Zehnder interferometer that directs the target photon in different output ports dependent on whether the control photon did interact with the QD or not. The pulse scheme and full operational principle of the CNOT gate are displayed in Fig.~\ref{fig:fig3}.b that also tracks the evolution of the quantum state of the whole system for the case where initially both control and target photons are in the state "1". Further details on the CNOT photon gate can be found in Supplementary Material. Importantly the basic functionalities of the protocol are experimentally feasible where the main considerations apart from the chiral spin interface are: i) the $\pi/2$ rotation of the spin qubit that can be achieved either optically \cite{Press2008Nature, Greilich2009NPHYS, Carter2013NPHOT} or with a microwave source \cite{Kroner2008PRL}, ii) a low loss optical interface between the chiral waveguide and a dielectric waveguide that can be readily engineered \cite{Mahmoodian2015InPrep}, and iii) reconfigurable optical circuits obtained from electromechanical transduction where the transmission of the beam splitter can be switched from 100/0 to 50/50 \cite{Akihama2011OE}. Based on these considerations it can be shown that the achievable entanglement fidelity of the CNOT gate is $96 \%$, which likely can be improved further by engineering GPWs with even higher $\beta_{\mathrm{dir}}.$

Finally, it is interesting to note that the photonic states in a GPW bear resemblance to the electronic edge states in quantum spin Hall systems, also known as 2D topological insulators \cite{Qi2011RMP}. In the latter case, the edge states are helical with the electron spin being perpendicular to the momentum that for a given propagation direction, the opposite direction carries the opposite spin. The same helicity is present in a GPW leading to a chiral coupling and thus unidirectional single-photon emission. The photonic-crystal platform allows engineering photonic materials at a length scale much below the lattice parameter allowing to tailor the dispersion and potentially therefore the topology of the structure. Furthermore, the excellent coherence properties of photons and availability of photon sources make photonics distinct from the case of electronics, and may lead to entirely new realms of studying topological effects.

A chiral photon-emitter interface represents a novel way of deterministically coupling single quanta of light and matter. It is expected to have widespread applications for scalable quantum-information processing utilizing photons. The general principle of engineering chiral interactions can very likely be extended to other platforms than considered here, for instance to the case of atoms in photonic-crystal structures \cite{Tiecke2014Nature, Goban2014NCOM}, nitrogen vacancy center in diamonds, or superconducting qubits \cite{Hoi2012PRL}. In the context of photon transport and scattering, chiral waveguide may be used to create novel topological states of photons that even could be exploited in a new regime of strong nonlinear photon interactions.

We thank Anders S{\o}rensen and Dirk Witthaut for valuable discussions and gratefully acknowledge financial support from the Villum Foundation, the Carlsberg Foundation, the Danish Council for Independent Research
(Natural Sciences and Technology and Production Sciences), and the European Research Council (ERC Consolidator Grant ALLQUANTUM). JDS and EHL acknowledge the support from the GRL and KIST institutional programs.

\section{Methods}
\emph{The sample.} The sample is grown by molecular beam epitaxy on an undoped (100) GaAs substrate. A 160-nm-thick GaAs membrane is grown on a 1.42-$\mu$m-thick Al$_{0.75}$Ga$_{0.25}$As sacrificial layer. The membrane contains a single layer of low-density self-assembled In(Ga)As quantum dots, emitting around 920 nm at 10K (density $\simeq$50 $\mu$m$^{-2}$). The photonic nanostructures are patterned by electron beam lithography at 100 keV on a 500-nm-thick electron beam resist (ZEP 520A) and aligned to the [011] or [0-11] directions of the GaAs substrate. The photonic-crystal holes are etched by an inductively coupled plasma in a BCl$_3$/Cl$_2$/Ar chemistry at 0 $^\circ$C. The remaining photoresist is stripped using hot N-Methyl-2-pyrrolidone and the membranes are subsequently released in a wet etching step, which removes the AlGaAs sacrificial layer using a hydrofluoric acid solution (10\% w/w).
\\

\emph{Experiment} The measurements are performed in a helium-bath cryostat operating at 4.2 K. The samples are mounted on a stack of piezoelectric nano-positioning stages allowing for precise positioning of the sample with respect to the objective. A superconducting coil is able to generate magnetic fields up to 9 T and surrounds the stage and sample such that the generated field points along the QD growth direction. The QDs are optically pumped by a  Ti:Sapphire laser at an emission wavelength of 842 nm. The laser is operated in continuous-wave mode for the spectral measurements and in pulsed mode (3-ps pulses at a repetition rate of 76 MHz) for lifetime and correlation measurements. Both excitation and collection are performed through the same microscope objective (NA $=$ 0.65). The excitation beam is moved to the center of the waveguide while the emission is collected from both gratings at the same time. The emission from the two gratings is coupled into separate single-mode polarization-maintaining fibers that lead to each their own spectrometer setup (1200 grooves/mm grating), where avalanche photo diodes are used for time resolved measurements, and a CCD is used to record the spectra.

\bibliography{bigBib}

\begin{thebibliography}{30}%
\makeatletter
\providecommand \@ifxundefined [1]{%
 \@ifx{#1\undefined}
}%
\providecommand \@ifnum [1]{%
 \ifnum #1\expandafter \@firstoftwo
 \else \expandafter \@secondoftwo
 \fi
}%
\providecommand \@ifx [1]{%
 \ifx #1\expandafter \@firstoftwo
 \else \expandafter \@secondoftwo
 \fi
}%
\providecommand \natexlab [1]{#1}%
\providecommand \enquote  [1]{``#1''}%
\providecommand \bibnamefont  [1]{#1}%
\providecommand \bibfnamefont [1]{#1}%
\providecommand \citenamefont [1]{#1}%
\providecommand \href@noop [0]{\@secondoftwo}%
\providecommand \href [0]{\begingroup \@sanitize@url \@href}%
\providecommand \@href[1]{\@@startlink{#1}\@@href}%
\providecommand \@@href[1]{\endgroup#1\@@endlink}%
\providecommand \@sanitize@url [0]{\catcode `\\12\catcode `\$12\catcode
  `\&12\catcode `\#12\catcode `\^12\catcode `\_12\catcode `\%12\relax}%
\providecommand \@@startlink[1]{}%
\providecommand \@@endlink[0]{}%
\providecommand \url  [0]{\begingroup\@sanitize@url \@url }%
\providecommand \@url [1]{\endgroup\@href {#1}{\urlprefix }}%
\providecommand \urlprefix  [0]{URL }%
\providecommand \Eprint [0]{\href }%
\providecommand \doibase [0]{http://dx.doi.org/}%
\providecommand \selectlanguage [0]{\@gobble}%
\providecommand \bibinfo  [0]{\@secondoftwo}%
\providecommand \bibfield  [0]{\@secondoftwo}%
\providecommand \translation [1]{[#1]}%
\providecommand \BibitemOpen [0]{}%
\providecommand \bibitemStop [0]{}%
\providecommand \bibitemNoStop [0]{.\EOS\space}%
\providecommand \EOS [0]{\spacefactor3000\relax}%
\providecommand \BibitemShut  [1]{\csname bibitem#1\endcsname}%
\let\auto@bib@innerbib\@empty
\bibitem [{\citenamefont {Mitsch}\ \emph {et~al.}(2014)\citenamefont {Mitsch},
  \citenamefont {Sayrin}, \citenamefont {Albrecht}, \citenamefont
  {Schneeweiss},\ and\ \citenamefont {Rauschenbeutel}}]{Mitsch2014arXiv}%
  \BibitemOpen
  \bibfield  {author} {\bibinfo {author} {\bibfnamefont {R.}~\bibnamefont
  {Mitsch}}, \bibinfo {author} {\bibfnamefont {C.}~\bibnamefont {Sayrin}},
  \bibinfo {author} {\bibfnamefont {B.}~\bibnamefont {Albrecht}}, \bibinfo
  {author} {\bibfnamefont {P.}~\bibnamefont {Schneeweiss}}, \ and\ \bibinfo
  {author} {\bibfnamefont {A.}~\bibnamefont {Rauschenbeutel}},\ }\href@noop {}
  {\bibfield  {journal} {\bibinfo  {journal} {arXiv:1406.0896}\ } (\bibinfo
  {year} {2014})}\BibitemShut {NoStop}%
\bibitem [{\citenamefont {Luxmoore}\ \emph {et~al.}(2013)\citenamefont
  {Luxmoore}, \citenamefont {Wasley}, \citenamefont {Ramsay}, \citenamefont
  {Thijssen}, \citenamefont {Oulton}, \citenamefont {Hugues}, \citenamefont
  {Kasture}, \citenamefont {Achanta}, \citenamefont {Fox},\ and\ \citenamefont
  {Skolnick}}]{Luxmoore2013PRL}%
  \BibitemOpen
  \bibfield  {author} {\bibinfo {author} {\bibfnamefont {I.~J.}\ \bibnamefont
  {Luxmoore}}, \bibinfo {author} {\bibfnamefont {N.~A.}\ \bibnamefont
  {Wasley}}, \bibinfo {author} {\bibfnamefont {A.~J.}\ \bibnamefont {Ramsay}},
  \bibinfo {author} {\bibfnamefont {A.~C.~T.}\ \bibnamefont {Thijssen}},
  \bibinfo {author} {\bibfnamefont {R.}~\bibnamefont {Oulton}}, \bibinfo
  {author} {\bibfnamefont {M.}~\bibnamefont {Hugues}}, \bibinfo {author}
  {\bibfnamefont {S.}~\bibnamefont {Kasture}}, \bibinfo {author} {\bibfnamefont
  {V.~G.}\ \bibnamefont {Achanta}}, \bibinfo {author} {\bibfnamefont {A.~M.}\
  \bibnamefont {Fox}}, \ and\ \bibinfo {author} {\bibfnamefont {M.~S.}\
  \bibnamefont {Skolnick}},\ }\href@noop {} {\bibfield  {journal} {\bibinfo
  {journal} {Phys. Rev. Lett.}\ }\textbf {\bibinfo {volume} {110}},\ \bibinfo
  {pages} {037402} (\bibinfo {year} {2013})}\BibitemShut {NoStop}%
\bibitem [{\citenamefont {Junge}\ \emph {et~al.}(2013)\citenamefont {Junge},
  \citenamefont {\protect{O'Shea}}, \citenamefont {Volz},\ and\ \citenamefont
  {Rauschenbeutel}}]{Junge2013PRL}%
  \BibitemOpen
  \bibfield  {author} {\bibinfo {author} {\bibfnamefont {C.}~\bibnamefont
  {Junge}}, \bibinfo {author} {\bibfnamefont {D.}~\bibnamefont
  {\protect{O'Shea}}}, \bibinfo {author} {\bibfnamefont {J.}~\bibnamefont
  {Volz}}, \ and\ \bibinfo {author} {\bibfnamefont {A.}~\bibnamefont
  {Rauschenbeutel}},\ }\href@noop {} {\bibfield  {journal} {\bibinfo  {journal}
  {Phys. Rev. Lett.}\ }\textbf {\bibinfo {volume} {110}},\ \bibinfo {pages}
  {213604} (\bibinfo {year} {2013})}\BibitemShut {NoStop}%
\bibitem [{\citenamefont {O'Brien}\ \emph {et~al.}(2009)\citenamefont
  {O'Brien}, \citenamefont {Furusawa},\ and\ \citenamefont
  {Vuckovic}}]{Obrien2009NPHOT}%
  \BibitemOpen
  \bibfield  {author} {\bibinfo {author} {\bibfnamefont {J.~L.}\ \bibnamefont
  {O'Brien}}, \bibinfo {author} {\bibfnamefont {A.}~\bibnamefont {Furusawa}}, \
  and\ \bibinfo {author} {\bibfnamefont {J.}~\bibnamefont {Vuckovic}},\
  }\href@noop {} {\bibfield  {journal} {\bibinfo  {journal} {Nat. Photonics}\
  }\textbf {\bibinfo {volume} {3}},\ \bibinfo {pages} {687} (\bibinfo {year}
  {2009})}\BibitemShut {NoStop}%
\bibitem [{\citenamefont {Stannigel}\ \emph {et~al.}(2012)\citenamefont
  {Stannigel}, \citenamefont {Rabl},\ and\ \citenamefont
  {Zoller}}]{Stannigel2012NJP}%
  \BibitemOpen
  \bibfield  {author} {\bibinfo {author} {\bibfnamefont {K.}~\bibnamefont
  {Stannigel}}, \bibinfo {author} {\bibfnamefont {P.}~\bibnamefont {Rabl}}, \
  and\ \bibinfo {author} {\bibfnamefont {P.}~\bibnamefont {Zoller}},\
  }\href@noop {} {\bibfield  {journal} {\bibinfo  {journal} {New J. Phys.}\
  }\textbf {\bibinfo {volume} {14}},\ \bibinfo {pages} {063014} (\bibinfo
  {year} {2012})}\BibitemShut {NoStop}%
\bibitem [{\citenamefont {Hafezi}\ \emph {et~al.}(2010)\citenamefont {Hafezi},
  \citenamefont {Demler}, \citenamefont {Lukin},\ and\ \citenamefont
  {Taylor}}]{Hafezi2011NPHYS}%
  \BibitemOpen
  \bibfield  {author} {\bibinfo {author} {\bibfnamefont {M.}~\bibnamefont
  {Hafezi}}, \bibinfo {author} {\bibfnamefont {E.~A.}\ \bibnamefont {Demler}},
  \bibinfo {author} {\bibfnamefont {M.~D.}\ \bibnamefont {Lukin}}, \ and\
  \bibinfo {author} {\bibfnamefont {J.~M.}\ \bibnamefont {Taylor}},\
  }\href@noop {} {\bibfield  {journal} {\bibinfo  {journal} {Nat. Phys.}\
  }\textbf {\bibinfo {volume} {7}},\ \bibinfo {pages} {907} (\bibinfo {year}
  {2010})}\BibitemShut {NoStop}%
\bibitem [{\citenamefont {Kraus}\ \emph {et~al.}(2012)\citenamefont {Kraus},
  \citenamefont {Lahini}, \citenamefont {Ringel}, \citenamefont {Verbin},\ and\
  \citenamefont {Zilberberg}}]{Kraus2012PRL}%
  \BibitemOpen
  \bibfield  {author} {\bibinfo {author} {\bibfnamefont {Y.~E.}\ \bibnamefont
  {Kraus}}, \bibinfo {author} {\bibfnamefont {Y.}~\bibnamefont {Lahini}},
  \bibinfo {author} {\bibfnamefont {Z.}~\bibnamefont {Ringel}}, \bibinfo
  {author} {\bibfnamefont {M.}~\bibnamefont {Verbin}}, \ and\ \bibinfo {author}
  {\bibfnamefont {O.}~\bibnamefont {Zilberberg}},\ }\href@noop {} {\bibfield
  {journal} {\bibinfo  {journal} {Phys. Rev. Lett.}\ }\textbf {\bibinfo
  {volume} {109}},\ \bibinfo {pages} {106402} (\bibinfo {year}
  {2012})}\BibitemShut {NoStop}%
\bibitem [{\citenamefont {\protect{Rodriguez-Fortu\~{n}o}}\ \emph
  {et~al.}(2013)\citenamefont {\protect{Rodriguez-Fortu\~{n}o}}, \citenamefont
  {Marino}, \citenamefont {Ginzburg}, \citenamefont {O'Connor}, \citenamefont
  {Martinez}, \citenamefont {Wurtz},\ and\ \citenamefont
  {Zayats}}]{Rodriguez2013Science}%
  \BibitemOpen
  \bibfield  {author} {\bibinfo {author} {\bibfnamefont {F.~J.}\ \bibnamefont
  {\protect{Rodriguez-Fortu\~{n}o}}}, \bibinfo {author} {\bibfnamefont
  {G.}~\bibnamefont {Marino}}, \bibinfo {author} {\bibfnamefont
  {P.}~\bibnamefont {Ginzburg}}, \bibinfo {author} {\bibfnamefont
  {D.}~\bibnamefont {O'Connor}}, \bibinfo {author} {\bibfnamefont
  {A.}~\bibnamefont {Martinez}}, \bibinfo {author} {\bibfnamefont {G.~A.}\
  \bibnamefont {Wurtz}}, \ and\ \bibinfo {author} {\bibfnamefont {A.~V.}\
  \bibnamefont {Zayats}},\ }\href@noop {} {\bibfield  {journal} {\bibinfo
  {journal} {Science}\ }\textbf {\bibinfo {volume} {340}},\ \bibinfo {pages}
  {328} (\bibinfo {year} {2013})}\BibitemShut {NoStop}%
\bibitem [{\citenamefont {Petersen}\ \emph {et~al.}(2014)\citenamefont
  {Petersen}, \citenamefont {Volz},\ and\ \citenamefont
  {Rauschenbeutel}}]{Petersen2014Science}%
  \BibitemOpen
  \bibfield  {author} {\bibinfo {author} {\bibfnamefont {J.}~\bibnamefont
  {Petersen}}, \bibinfo {author} {\bibfnamefont {J.}~\bibnamefont {Volz}}, \
  and\ \bibinfo {author} {\bibfnamefont {A.}~\bibnamefont {Rauschenbeutel}},\
  }\href@noop {} {\bibfield  {journal} {\bibinfo  {journal} {Science}\ }\textbf
  {\bibinfo {volume} {346}},\ \bibinfo {pages} {67} (\bibinfo {year}
  {2014})}\BibitemShut {NoStop}%
\bibitem [{\citenamefont {Neugebauer}\ \emph {et~al.}(2014)\citenamefont
  {Neugebauer}, \citenamefont {Bauer}, \citenamefont {Banzer},\ and\
  \citenamefont {Leuchs}}]{Neugebauer2014NL}%
  \BibitemOpen
  \bibfield  {author} {\bibinfo {author} {\bibfnamefont {M.}~\bibnamefont
  {Neugebauer}}, \bibinfo {author} {\bibfnamefont {T.}~\bibnamefont {Bauer}},
  \bibinfo {author} {\bibfnamefont {P.}~\bibnamefont {Banzer}}, \ and\ \bibinfo
  {author} {\bibfnamefont {G.}~\bibnamefont {Leuchs}},\ }\href@noop {}
  {\bibfield  {journal} {\bibinfo  {journal} {Nano Lett.}\ }\textbf {\bibinfo
  {volume} {14}},\ \bibinfo {pages} {2546} (\bibinfo {year}
  {2014})}\BibitemShut {NoStop}%
\bibitem [{\citenamefont {Lodahl}\ \emph {et~al.}(2013)\citenamefont {Lodahl},
  \citenamefont {Mahmoodian},\ and\ \citenamefont
  {Stobbe}}]{Lodahl2013arXivRMP}%
  \BibitemOpen
  \bibfield  {author} {\bibinfo {author} {\bibfnamefont {P.}~\bibnamefont
  {Lodahl}}, \bibinfo {author} {\bibfnamefont {S.}~\bibnamefont {Mahmoodian}},
  \ and\ \bibinfo {author} {\bibfnamefont {S.}~\bibnamefont {Stobbe}},\
  }\href@noop {} {\bibfield  {journal} {\bibinfo  {journal} {arXiv:1312.1079}\
  } (\bibinfo {year} {2013})}\BibitemShut {NoStop}%
\bibitem [{\citenamefont {Arcari}\ \emph {et~al.}(2014)\citenamefont {Arcari},
  \citenamefont {S\"ollner}, \citenamefont {Javadi}, \citenamefont
  {Lindskov~Hansen}, \citenamefont {Mahmoodian}, \citenamefont {Liu},
  \citenamefont {Thyrrestrup}, \citenamefont {Lee}, \citenamefont {Song},
  \citenamefont {Stobbe},\ and\ \citenamefont {Lodahl}}]{Arcari2014PRL}%
  \BibitemOpen
  \bibfield  {author} {\bibinfo {author} {\bibfnamefont {M.}~\bibnamefont
  {Arcari}}, \bibinfo {author} {\bibfnamefont {I.}~\bibnamefont {S\"ollner}},
  \bibinfo {author} {\bibfnamefont {A.}~\bibnamefont {Javadi}}, \bibinfo
  {author} {\bibfnamefont {S.}~\bibnamefont {Lindskov~Hansen}}, \bibinfo
  {author} {\bibfnamefont {S.}~\bibnamefont {Mahmoodian}}, \bibinfo {author}
  {\bibfnamefont {J.}~\bibnamefont {Liu}}, \bibinfo {author} {\bibfnamefont
  {H.}~\bibnamefont {Thyrrestrup}}, \bibinfo {author} {\bibfnamefont {E.~H.}\
  \bibnamefont {Lee}}, \bibinfo {author} {\bibfnamefont {J.~D.}\ \bibnamefont
  {Song}}, \bibinfo {author} {\bibfnamefont {S.}~\bibnamefont {Stobbe}}, \ and\
  \bibinfo {author} {\bibfnamefont {P.}~\bibnamefont {Lodahl}},\ }\href@noop {}
  {\bibfield  {journal} {\bibinfo  {journal} {Phys. Rev. Lett.}\ }\textbf
  {\bibinfo {volume} {113}},\ \bibinfo {pages} {093603} (\bibinfo {year}
  {2014})}\BibitemShut {NoStop}%
\bibitem [{\citenamefont {Bayer}\ \emph {et~al.}(2002)\citenamefont {Bayer},
  \citenamefont {Ortner}, \citenamefont {Stern}, \citenamefont {Kuther},
  \citenamefont {Gorbunov}, \citenamefont {Forchel}, \citenamefont {Hawrylak},
  \citenamefont {Fafard}, \citenamefont {Hinzer}, \citenamefont {Reinecke},
  \citenamefont {Walck}, \citenamefont {Reithmaier}, \citenamefont {Klopf},\
  and\ \citenamefont {Sch\"afer}}]{Bayer2002PRBb}%
  \BibitemOpen
  \bibfield  {author} {\bibinfo {author} {\bibfnamefont {M.}~\bibnamefont
  {Bayer}}, \bibinfo {author} {\bibfnamefont {G.}~\bibnamefont {Ortner}},
  \bibinfo {author} {\bibfnamefont {O.}~\bibnamefont {Stern}}, \bibinfo
  {author} {\bibfnamefont {A.}~\bibnamefont {Kuther}}, \bibinfo {author}
  {\bibfnamefont {A.~A.}\ \bibnamefont {Gorbunov}}, \bibinfo {author}
  {\bibfnamefont {A.}~\bibnamefont {Forchel}}, \bibinfo {author} {\bibfnamefont
  {P.}~\bibnamefont {Hawrylak}}, \bibinfo {author} {\bibfnamefont
  {S.}~\bibnamefont {Fafard}}, \bibinfo {author} {\bibfnamefont
  {K.}~\bibnamefont {Hinzer}}, \bibinfo {author} {\bibfnamefont {T.~L.}\
  \bibnamefont {Reinecke}}, \bibinfo {author} {\bibfnamefont {S.~N.}\
  \bibnamefont {Walck}}, \bibinfo {author} {\bibfnamefont {J.~P.}\ \bibnamefont
  {Reithmaier}}, \bibinfo {author} {\bibfnamefont {F.}~\bibnamefont {Klopf}}, \
  and\ \bibinfo {author} {\bibfnamefont {F.}~\bibnamefont {Sch\"afer}},\
  }\href@noop {} {\bibfield  {journal} {\bibinfo  {journal} {Phys. Rev. B}\
  }\textbf {\bibinfo {volume} {65}},\ \bibinfo {pages} {195315} (\bibinfo
  {year} {2002})}\BibitemShut {NoStop}%
\bibitem [{\citenamefont {Santori}\ \emph {et~al.}(2010)\citenamefont
  {Santori}, \citenamefont {Fattal},\ and\ \citenamefont
  {Yamamoto}}]{SantoriBook}%
  \BibitemOpen
  \bibfield  {author} {\bibinfo {author} {\bibfnamefont {C.}~\bibnamefont
  {Santori}}, \bibinfo {author} {\bibfnamefont {D.}~\bibnamefont {Fattal}}, \
  and\ \bibinfo {author} {\bibfnamefont {Y.}~\bibnamefont {Yamamoto}},\
  }\href@noop {} {\emph {\bibinfo {title} {Single-photon Devices and
  Applications}}}\ (\bibinfo  {publisher} {John Wiley \& Sons, Inc.},\ \bibinfo
  {year} {2010})\BibitemShut {NoStop}%
\bibitem [{\citenamefont {Nielsen}\ and\ \citenamefont
  {Chuang}(2010)}]{Nielsen2010Book}%
  \BibitemOpen
  \bibfield  {author} {\bibinfo {author} {\bibfnamefont {M.~A.}\ \bibnamefont
  {Nielsen}}\ and\ \bibinfo {author} {\bibfnamefont {I.~L.}\ \bibnamefont
  {Chuang}},\ }\href@noop {} {\emph {\bibinfo {title} {Quantum computation and
  quantum information}}}\ (\bibinfo  {publisher} {Cambridge university press},\
  \bibinfo {year} {2010})\BibitemShut {NoStop}%
\bibitem [{\citenamefont {Duan}\ and\ \citenamefont
  {Kimble}(2004)}]{Duan2004PRL}%
  \BibitemOpen
  \bibfield  {author} {\bibinfo {author} {\bibfnamefont {L.~M.}\ \bibnamefont
  {Duan}}\ and\ \bibinfo {author} {\bibfnamefont {H.~J.}\ \bibnamefont
  {Kimble}},\ }\href@noop {} {\bibfield  {journal} {\bibinfo  {journal} {Phys.
  Rev. Lett.}\ }\textbf {\bibinfo {volume} {92}},\ \bibinfo {pages} {127902}
  (\bibinfo {year} {2004})}\BibitemShut {NoStop}%
\bibitem [{\citenamefont {Chang}\ \emph {et~al.}(2008)\citenamefont {Chang},
  \citenamefont {Gritsev}, \citenamefont {Morigi}, \citenamefont {Vuleti\'{c}},
  \citenamefont {Lukin},\ and\ \citenamefont {Demler}}]{Chang2008NPHYS}%
  \BibitemOpen
  \bibfield  {author} {\bibinfo {author} {\bibfnamefont {D.~E.}\ \bibnamefont
  {Chang}}, \bibinfo {author} {\bibfnamefont {V.}~\bibnamefont {Gritsev}},
  \bibinfo {author} {\bibfnamefont {G.}~\bibnamefont {Morigi}}, \bibinfo
  {author} {\bibfnamefont {V.}~\bibnamefont {Vuleti\'{c}}}, \bibinfo {author}
  {\bibfnamefont {M.~D.}\ \bibnamefont {Lukin}}, \ and\ \bibinfo {author}
  {\bibfnamefont {E.~A.}\ \bibnamefont {Demler}},\ }\href@noop {} {\bibfield
  {journal} {\bibinfo  {journal} {Nat. Phys.}\ }\textbf {\bibinfo {volume}
  {4}},\ \bibinfo {pages} {884} (\bibinfo {year} {2008})}\BibitemShut {NoStop}%
\bibitem [{\citenamefont {Shen}\ and\ \citenamefont {Fan}(2005)}]{Shen2005OL}%
  \BibitemOpen
  \bibfield  {author} {\bibinfo {author} {\bibfnamefont {J.~T.}\ \bibnamefont
  {Shen}}\ and\ \bibinfo {author} {\bibfnamefont {S.}~\bibnamefont {Fan}},\
  }\href@noop {} {\bibfield  {journal} {\bibinfo  {journal} {Opt. Lett.}\
  }\textbf {\bibinfo {volume} {30}},\ \bibinfo {pages} {2001} (\bibinfo {year}
  {2005})}\BibitemShut {NoStop}%
\bibitem [{Note1()}]{Note1}%
  \BibitemOpen
  \bibinfo {note} {We note that the acquired transmission coefficient is at
  variance with the results reported in \cite {Shen2011PRL}. This mistake has
  been communicated to and acknowledged by the authors of the mentioned
  reference.}\BibitemShut {Stop}%
\bibitem [{\citenamefont {Press}\ \emph {et~al.}(2008)\citenamefont {Press},
  \citenamefont {Ladd}, \citenamefont {Zhang},\ and\ \citenamefont
  {Yamamoto}}]{Press2008Nature}%
  \BibitemOpen
  \bibfield  {author} {\bibinfo {author} {\bibfnamefont {D.}~\bibnamefont
  {Press}}, \bibinfo {author} {\bibfnamefont {T.~D.}\ \bibnamefont {Ladd}},
  \bibinfo {author} {\bibfnamefont {B.}~\bibnamefont {Zhang}}, \ and\ \bibinfo
  {author} {\bibfnamefont {Y.}~\bibnamefont {Yamamoto}},\ }\href@noop {}
  {\bibfield  {journal} {\bibinfo  {journal} {Nature}\ }\textbf {\bibinfo
  {volume} {456}},\ \bibinfo {pages} {218} (\bibinfo {year}
  {2008})}\BibitemShut {NoStop}%
\bibitem [{\citenamefont {Greilich}\ \emph {et~al.}(2009)\citenamefont
  {Greilich}, \citenamefont {Economou}, \citenamefont {Spatzek}, \citenamefont
  {Yakovlev}, \citenamefont {Reuter}, \citenamefont {Wieck}, \citenamefont
  {Reinecke},\ and\ \citenamefont {Bayer}}]{Greilich2009NPHYS}%
  \BibitemOpen
  \bibfield  {author} {\bibinfo {author} {\bibfnamefont {A.}~\bibnamefont
  {Greilich}}, \bibinfo {author} {\bibfnamefont {S.~E.}\ \bibnamefont
  {Economou}}, \bibinfo {author} {\bibfnamefont {S.}~\bibnamefont {Spatzek}},
  \bibinfo {author} {\bibfnamefont {D.~R.}\ \bibnamefont {Yakovlev}}, \bibinfo
  {author} {\bibfnamefont {D.}~\bibnamefont {Reuter}}, \bibinfo {author}
  {\bibfnamefont {A.~D.}\ \bibnamefont {Wieck}}, \bibinfo {author}
  {\bibfnamefont {T.~L.}\ \bibnamefont {Reinecke}}, \ and\ \bibinfo {author}
  {\bibfnamefont {M.}~\bibnamefont {Bayer}},\ }\href@noop {} {\bibfield
  {journal} {\bibinfo  {journal} {Nat. Phys.}\ }\textbf {\bibinfo {volume}
  {262}},\ \bibinfo {pages} {262} (\bibinfo {year} {2009})}\BibitemShut
  {NoStop}%
\bibitem [{\citenamefont {Carter}\ \emph {et~al.}(2013)\citenamefont {Carter},
  \citenamefont {Sweeney}, \citenamefont {Kim}, \citenamefont {Kim},
  \citenamefont {Solenov}, \citenamefont {Economou}, \citenamefont {Reinecke},
  \citenamefont {Yang}, \citenamefont {Bracker},\ and\ \citenamefont
  {Gammon}}]{Carter2013NPHOT}%
  \BibitemOpen
  \bibfield  {author} {\bibinfo {author} {\bibfnamefont {S.~G.}\ \bibnamefont
  {Carter}}, \bibinfo {author} {\bibfnamefont {T.~M.}\ \bibnamefont {Sweeney}},
  \bibinfo {author} {\bibfnamefont {M.}~\bibnamefont {Kim}}, \bibinfo {author}
  {\bibfnamefont {C.~S.}\ \bibnamefont {Kim}}, \bibinfo {author} {\bibfnamefont
  {D.}~\bibnamefont {Solenov}}, \bibinfo {author} {\bibfnamefont {S.~E.}\
  \bibnamefont {Economou}}, \bibinfo {author} {\bibfnamefont {T.~L.}\
  \bibnamefont {Reinecke}}, \bibinfo {author} {\bibfnamefont {L.}~\bibnamefont
  {Yang}}, \bibinfo {author} {\bibfnamefont {A.~S.}\ \bibnamefont {Bracker}}, \
  and\ \bibinfo {author} {\bibfnamefont {D.}~\bibnamefont {Gammon}},\
  }\href@noop {} {\bibfield  {journal} {\bibinfo  {journal} {Nat. Photonics}\
  }\textbf {\bibinfo {volume} {7}},\ \bibinfo {pages} {329} (\bibinfo {year}
  {2013})}\BibitemShut {NoStop}%
\bibitem [{\citenamefont {Kroner}\ \emph {et~al.}(2008)\citenamefont {Kroner},
  \citenamefont {Weiss}, \citenamefont {Biedermann}, \citenamefont {Seidl},
  \citenamefont {Manus}, \citenamefont {Holleitner}, \citenamefont {Badolato},
  \citenamefont {Petroff}, \citenamefont {Gerardot}, \citenamefont
  {Warburton},\ and\ \citenamefont {Karrai}}]{Kroner2008PRL}%
  \BibitemOpen
  \bibfield  {author} {\bibinfo {author} {\bibfnamefont {M.}~\bibnamefont
  {Kroner}}, \bibinfo {author} {\bibfnamefont {K.~M.}\ \bibnamefont {Weiss}},
  \bibinfo {author} {\bibfnamefont {B.}~\bibnamefont {Biedermann}}, \bibinfo
  {author} {\bibfnamefont {S.}~\bibnamefont {Seidl}}, \bibinfo {author}
  {\bibfnamefont {S.}~\bibnamefont {Manus}}, \bibinfo {author} {\bibfnamefont
  {A.~W.}\ \bibnamefont {Holleitner}}, \bibinfo {author} {\bibfnamefont
  {A.}~\bibnamefont {Badolato}}, \bibinfo {author} {\bibfnamefont {P.~M.}\
  \bibnamefont {Petroff}}, \bibinfo {author} {\bibfnamefont {B.~D.}\
  \bibnamefont {Gerardot}}, \bibinfo {author} {\bibfnamefont {R.~J.}\
  \bibnamefont {Warburton}}, \ and\ \bibinfo {author} {\bibfnamefont
  {K.}~\bibnamefont {Karrai}},\ }\href@noop {} {\bibfield  {journal} {\bibinfo
  {journal} {Phys. Rev. Lett.}\ }\textbf {\bibinfo {volume} {100}},\ \bibinfo
  {pages} {156803} (\bibinfo {year} {2008})}\BibitemShut {NoStop}%
\bibitem [{\citenamefont {Mahmoodian}\ and\ \citenamefont {{\it et
  al.}}(2015)}]{Mahmoodian2015InPrep}%
  \BibitemOpen
  \bibfield  {author} {\bibinfo {author} {\bibfnamefont {S.}~\bibnamefont
  {Mahmoodian}}\ and\ \bibinfo {author} {\bibnamefont {{\it et al.}}},\
  }\href@noop {} {\bibfield  {journal} {\bibinfo  {journal} {In preparation}\ }
  (\bibinfo {year} {2015})}\BibitemShut {NoStop}%
\bibitem [{\citenamefont {Akihama}\ \emph {et~al.}(2011)\citenamefont
  {Akihama}, \citenamefont {Kanamori},\ and\ \citenamefont
  {Hane}}]{Akihama2011OE}%
  \BibitemOpen
  \bibfield  {author} {\bibinfo {author} {\bibfnamefont {Y.}~\bibnamefont
  {Akihama}}, \bibinfo {author} {\bibfnamefont {Y.}~\bibnamefont {Kanamori}}, \
  and\ \bibinfo {author} {\bibfnamefont {K.}~\bibnamefont {Hane}},\ }\href@noop
  {} {\bibfield  {journal} {\bibinfo  {journal} {Opt. Express}\ }\textbf
  {\bibinfo {volume} {19}},\ \bibinfo {pages} {23658} (\bibinfo {year}
  {2011})}\BibitemShut {NoStop}%
\bibitem [{\citenamefont {Qi}\ and\ \citenamefont {Zhang}(2011)}]{Qi2011RMP}%
  \BibitemOpen
  \bibfield  {author} {\bibinfo {author} {\bibfnamefont {X.-L.}\ \bibnamefont
  {Qi}}\ and\ \bibinfo {author} {\bibfnamefont {S.-C.}\ \bibnamefont {Zhang}},\
  }\href@noop {} {\bibfield  {journal} {\bibinfo  {journal} {Rev. Mod. Phys.}\
  }\textbf {\bibinfo {volume} {83}},\ \bibinfo {pages} {1057} (\bibinfo {year}
  {2011})}\BibitemShut {NoStop}%
\bibitem [{\citenamefont {Tiecke}\ \emph {et~al.}(2014)\citenamefont {Tiecke},
  \citenamefont {Thompson}, \citenamefont {de~Leon}, \citenamefont {Liu},
  \citenamefont {Vuleti\`{c}},\ and\ \citenamefont {Lukin}}]{Tiecke2014Nature}%
  \BibitemOpen
  \bibfield  {author} {\bibinfo {author} {\bibfnamefont {T.~G.}\ \bibnamefont
  {Tiecke}}, \bibinfo {author} {\bibfnamefont {J.~D.}\ \bibnamefont
  {Thompson}}, \bibinfo {author} {\bibfnamefont {N.~P.}\ \bibnamefont
  {de~Leon}}, \bibinfo {author} {\bibfnamefont {L.~R.}\ \bibnamefont {Liu}},
  \bibinfo {author} {\bibfnamefont {V.}~\bibnamefont {Vuleti\`{c}}}, \ and\
  \bibinfo {author} {\bibfnamefont {M.~D.}\ \bibnamefont {Lukin}},\ }\href@noop
  {} {\bibfield  {journal} {\bibinfo  {journal} {Nature}\ }\textbf {\bibinfo
  {volume} {508}},\ \bibinfo {pages} {241} (\bibinfo {year}
  {2014})}\BibitemShut {NoStop}%
\bibitem [{\citenamefont {Goban}\ \emph {et~al.}(2014)\citenamefont {Goban},
  \citenamefont {Hung}, \citenamefont {Yu}, \citenamefont {Hood}, \citenamefont
  {Muniz}, \citenamefont {Lee}, \citenamefont {Martin}, \citenamefont
  {McClung}, \citenamefont {Choi}, \citenamefont {Chang}, \citenamefont
  {Painter},\ and\ \citenamefont {Kimble}}]{Goban2014NCOM}%
  \BibitemOpen
  \bibfield  {author} {\bibinfo {author} {\bibfnamefont {A.}~\bibnamefont
  {Goban}}, \bibinfo {author} {\bibfnamefont {C.~L.}\ \bibnamefont {Hung}},
  \bibinfo {author} {\bibfnamefont {S.~P.}\ \bibnamefont {Yu}}, \bibinfo
  {author} {\bibfnamefont {J.~D.}\ \bibnamefont {Hood}}, \bibinfo {author}
  {\bibfnamefont {J.~A.}\ \bibnamefont {Muniz}}, \bibinfo {author}
  {\bibfnamefont {J.~H.}\ \bibnamefont {Lee}}, \bibinfo {author} {\bibfnamefont
  {M.~J.}\ \bibnamefont {Martin}}, \bibinfo {author} {\bibfnamefont {A.~C.}\
  \bibnamefont {McClung}}, \bibinfo {author} {\bibfnamefont {K.~S.}\
  \bibnamefont {Choi}}, \bibinfo {author} {\bibfnamefont {D.~E.}\ \bibnamefont
  {Chang}}, \bibinfo {author} {\bibfnamefont {O.}~\bibnamefont {Painter}}, \
  and\ \bibinfo {author} {\bibfnamefont {H.~J.}\ \bibnamefont {Kimble}},\
  }\href@noop {} {\bibfield  {journal} {\bibinfo  {journal} {Nat. Commun.}\
  }\textbf {\bibinfo {volume} {5}},\ \bibinfo {pages} {4808} (\bibinfo {year}
  {2014})}\BibitemShut {NoStop}%
\bibitem [{\citenamefont {Hoi}\ \emph {et~al.}(2012)\citenamefont {Hoi},
  \citenamefont {Palomaki}, \citenamefont {Lindkvist}, \citenamefont
  {Johansson}, \citenamefont {Delsing},\ and\ \citenamefont
  {Wilson}}]{Hoi2012PRL}%
  \BibitemOpen
  \bibfield  {author} {\bibinfo {author} {\bibfnamefont {I.-C.}\ \bibnamefont
  {Hoi}}, \bibinfo {author} {\bibfnamefont {T.}~\bibnamefont {Palomaki}},
  \bibinfo {author} {\bibfnamefont {J.}~\bibnamefont {Lindkvist}}, \bibinfo
  {author} {\bibfnamefont {G.}~\bibnamefont {Johansson}}, \bibinfo {author}
  {\bibfnamefont {P.}~\bibnamefont {Delsing}}, \ and\ \bibinfo {author}
  {\bibfnamefont {C.~M.}\ \bibnamefont {Wilson}},\ }\href@noop {} {\bibfield
  {journal} {\bibinfo  {journal} {Phys. Rev. Lett.}\ }\textbf {\bibinfo
  {volume} {108}},\ \bibinfo {pages} {263601} (\bibinfo {year}
  {2012})}\BibitemShut {NoStop}%
\bibitem [{\citenamefont {Shen}\ \emph {et~al.}(2011)\citenamefont {Shen},
  \citenamefont {Bradford},\ and\ \citenamefont {Shen}}]{Shen2011PRL}%
  \BibitemOpen
  \bibfield  {author} {\bibinfo {author} {\bibfnamefont {Y.}~\bibnamefont
  {Shen}}, \bibinfo {author} {\bibfnamefont {M.}~\bibnamefont {Bradford}}, \
  and\ \bibinfo {author} {\bibfnamefont {J.-T.}\ \bibnamefont {Shen}},\
  }\href@noop {} {\bibfield  {journal} {\bibinfo  {journal} {Phys. Rev. Lett.}\
  }\textbf {\bibinfo {volume} {107}},\ \bibinfo {pages} {173902} (\bibinfo
  {year} {2011})}\BibitemShut {NoStop}%
\end{thebibliography}%

\clearpage
\newpage
\onecolumngrid

\section{Supplementary material}

\subsection{Sample design}

\begin{figure*}[th!]
\begin{center}
\includegraphics[width=13cm]{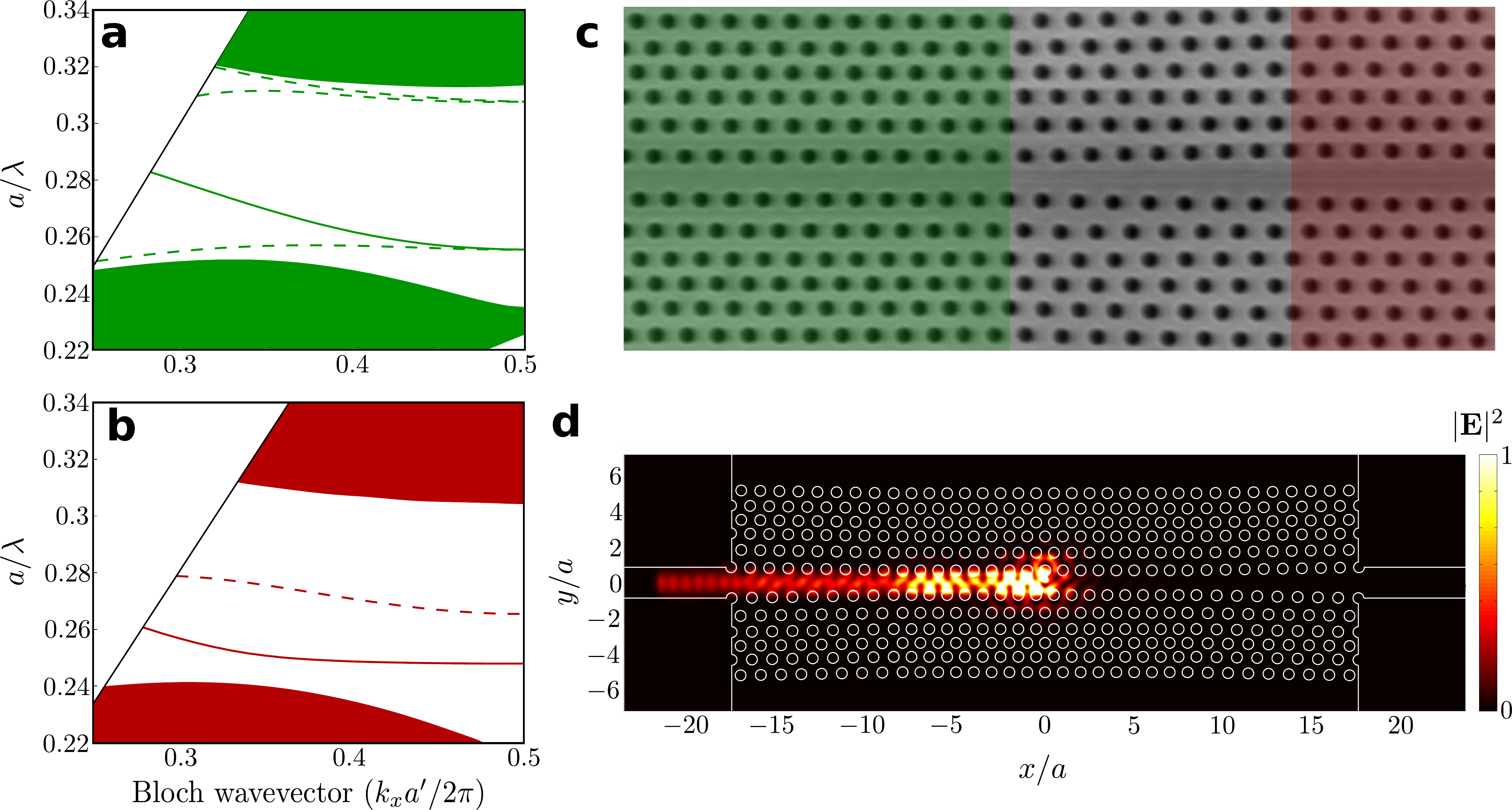}
\end{center}
\caption{ {\bf a)} Band diagram of the dispersion engineered GPW with period $a$, refractive index $n=3.4638$, hole radius
$r=0.3 a$, and membrane thickness $t=2a/3$. The solid line is the band of interest, while broken lines and solid shading represent other bands and continuum states respectively. The black line indicates the light line. {\bf b)} Band diagram of the out-coupling section of the device. The parameters are the same as in a), but the lattice constant is $a'=1.07 a$ and the waveguide is up-down symmetric. {\bf c)} Electron-microscope image of sample with the glide-plane section of waveguide highlighted in green and the OCW section highlighted in red. An adiabatic transition region joins the GPW and OCW regions. {\bf d)} Finite element calculation of the electric field intensity (saturated colour scale in arbitrary units) emitted by a right hand circularly polarized dipole with frequency $a/\lambda=0.26$.} \label{fig:transition} \end{figure*}

Figure \ref{fig:transition}.a shows the band diagram of the glide-plane waveguide (GPW) and
Fig.~\ref{fig:transition}.b the band diagram of the out-coupling waveguide (OCW). The OCW is designed such that it has a mode with a low group index $(n_g)$ at the frequency of operation of the GPW $(a/\lambda \sim 0.26)$. In our sample, the out-coupling waveguide is coupled to a ridge waveguide terminated by an out-coupling grating. Figure \ref{fig:transition}.c shows an SEM of the sample with a GPW (green highlight), a symmetric waveguide (red
highlight) and an adiabatic taper (grey). The two waveguides have different symmetries and cannot be directly coupled to each other,
therefore we designed the adiabatic taper to efficiently couple the waveguide modes together. Figure \ref{fig:transition}.d shows an $x,y$ slice ($z=0$) of the emitted field intensity of a right hand circularly polarized (RHCP) dipole ($\hat{\mathbf e}_x + i \hat{\mathbf e}_y$) computed using finite element software. The dipole is positioned at a point where the left propagating mode of the waveguide is RHCP while the right propagating mode has the orthogonal polarization. The finite element calculation shows that the adiabatic taper couples the mode efficiently to the symmetric OCW with very little back reflection.  The computation domain is surrounded by perfectly-matched layers to absorb the out-going field. From this simulation we extracted a directionality of $F_\textrm{dir} = 98 \%$ with an efficiency of $\sim 85\%$, i.e., $85\%$ of the light emitted from the dipole is coupled to the ridge waveguide on the left. Since we expect that the quantum dot couples photons with near-unity efficiency into the GPW, most of the losses are likely due to out-of-plane scattering at the adiabatic transition and at the interface to the ridge waveguide. This can be improved by further optimizing the adiabatic taper section.

\subsection{Determining the directionality}

 The directionality of the photon emission can be extracted from the experiments using spectra like those presented in Fig.~\ref{fig:fig2}. For varying magnetic field, both the left and right-propagating modes are collected and the emission lines in the spectra are fitted by Lorentzians. The full-width-half-maxima of the fitted Lorentzians are used as the bin size and the total number of events centered at the peak are summed. These integrated count rates are used to extract the directionality for photons coupled out either left (L) or right (R) from the GPW (cf. data in Fig.~\ref{fig:fig2}.b):
\begin{equation}\label{fdirL}
F_{\textrm{dir,L/R}}={I_{+/-,\textrm{L/R}} \over I_{-,\textrm{L/R}}+I_{+,\textrm{L/R}}}
\end{equation}
where $I_{-,\textrm{R/L}}$ and $I_{+,\textrm{R/L}}$  denote the integrated count rates for the $\sigma_-$ and $\sigma_+$ transitions into the directions denoted by the subscript, respectively. Here we have assumed that both dipole transitions are populated equally at the excitation.
Figure \ref{fig:fig2}.d displays $F_{\textrm{dir}}=\left( F_{\textrm{dir,L}}+F_{\textrm{dir,R}} \right) / 2$  extracted for QD A and B and as a function of applied magnetic field.

The directional $\beta$-factor is defined as
\begin{equation}
\beta_\textrm{dir} = \frac{\max\left[{\Gamma_{\textrm R}, \Gamma_{\textrm L}}\right]}{\Gamma_\textrm{wg}+\gamma_\textrm{rad}},
\end{equation}
where $\Gamma_{\textrm R}$ and $\Gamma_{\textrm L}$ are the decay rates for coupling to right and left propagating waveguide modes, respectively, $ \Gamma_\textrm{wg} = \Gamma_\textrm{L} + \Gamma_\textrm{R}$, and $\gamma_{\textrm{rad}}$ is the coupling to non-guided radiation modes. Therefore $\beta_\textrm{dir}$ quantifies the fraction of emission into a waveguide mode propagating in a single direction as opposed to all other optical modes. This can also be written $\beta_\textrm{dir} = \beta F_\textrm{dir}$, where $\beta = \Gamma_\textrm{wg}/(\Gamma_\textrm{wg} + \gamma_\textrm{rad})$ and $F_\textrm{dir} = \max\left[{\Gamma_{\textrm R}, \Gamma_{\textrm L}}\right] /\Gamma_\textrm{wg}$.

\subsection{The photon-photon CNOT gate}

A photon-photon CNOT gate requires a single control photon to flip the state of a target photon. In our scheme the quantum states of the photons are encoded in spatially separated waveguide modes, and thus by manipulating the internal state of a quantum dot, the control photon determines the path of the target photon. The external magnetic field $B_{\textrm{ext}}$ splits the transitions and decreases the rate of diagonal transitions. The CNOT gate requires control and target photons to interact sequentially with the QD. This can be achieved using a reconfigurable beam splitter. The distance between the two arms of the beam s½plitter is reconfigured to alternate between no coupling and a 50:50 splitting ratio. The reconfigurable beam splitter allows mapping the spatial modes $|1\rangle_\textrm{c}$ and $|0 \rangle_\textrm{t}$ to distinct temporal modes. The red photon passes through the circuit first, the beam splitters are then reconfigured to 50:50 coupling, and then the blue photon enters the circuit. We note that because the QD optical transitions have orthogonal circular polarization, the control and target photons propagate in opposite directions.

The following sequence is used to implement the CNOT gate:
\begin{enumerate}
        \item The emitter is initialized in the state $\ket{\uparrow}$.
        \item A ${\pi\over 2}$-rotation transforms the spin state to ${\frac{1}{\sqrt{2}}}\left(\ket{\uparrow}+\ket{\downarrow} \right)$.
        \item If there is a control photon in the $| 1 \rangle_\textrm{c}$ state,  it scatters off the red transition and causes a $\pi$ phase shift  between the two spin components. If the photon is in the $| 0 \rangle_\textrm{c}$ state, the QD remains in the state ${\frac{1}{\sqrt{2}}}\left(\ket{\uparrow}+\ket{\downarrow} \right)$.
        \item A $-{\pi\over 2}$-rotation of the spin maps ${\frac{1}{\sqrt{2}}}\left(\ket{\uparrow}-\ket{\downarrow} \right)$ to $\ket{\downarrow}$, but maps ${\frac{1}{\sqrt{2}}}\left(\ket{\uparrow}+\ket{\downarrow} \right)$ to  the original state state $\ket{\uparrow}$.
        \item A single target photon in the state $| 0 \rangle_\textrm{t}$ ($| 1 \rangle_\textrm{t}$) exits in the state $| 1 \rangle_\textrm{t}$ ($| 0 \rangle_\textrm{t}$) if the control photon was in state $| 1 \rangle_\textrm{c}$, otherwise the state of the target photon remains unchanged.
        \item A quantum eraser protocol is employed: a $\frac{\pi}{2}$ rotation on the QD spin is followed by a measurement of the spin state using a coherent state. A $\pi$ phase is imparted on state $| 1 \rangle_\textrm{c}$ on the condition of measuring $\ket{\downarrow}$.
\end{enumerate}

We note that the final step is required to disentangle the spin state from the optical state, which occurs for a general input state. The measurement can be done by reading out the spin state using a weak coherent state, which introduces no additional errors. Here we present the fidelity ${\cal F}$ of an entanglement process and the minimum fidelity ${\cal F}_{\textrm{min}}$ for the gate. For an ideal CNOT gate the input state $(\ket{0_c}+\ket{1_c})\ket{0_t}/\sqrt{2}$ produces the maximally entangled Bell state $\ket{\Phi^+}$ as an output. The modulus square of the overlap of the output state $\ket{\textrm{out}}$ with the output of the ideal gate gives the fidelity ${\cal F} =|\langle \Phi^+\ket{\textrm{out}}|^2 = \beta_{\textrm{dir}}^2$. We have shown that values of $\beta_{\textrm{dir}}\sim 0.98$  can be achieved simultaneously for both transitions in our PCW corresponding to ${\cal F} = 0.96$ for an entanglement process. The gate operates at its lowest fidelity for the input state $\ket{0_c}(i\ket{0_t}+\ket{1_t})/\sqrt{2}$, giving ${\cal F}_{\textrm{min}} = (1-2\beta_{\textrm{dir}})^2$. For narrow-band pulses with $\beta_{\textrm{dir}}=0.98$ this gives a minimum fidelity ${\cal F}_{\textrm{min}} = 0.92$.

\end{document}